\documentclass[twocolumn,apj,numberedappendix,twocolappendix]{openjournal}
\usepackage{amsmath}
\usepackage{amssymb}
\usepackage{graphicx}
\usepackage{natbib}
\usepackage{orcidlink}
\usepackage{hyperref}

\usepackage{xcolor}

\definecolor{linkblue}{RGB}{0,70,140}

\hypersetup{
  colorlinks=true,
  linkcolor=linkblue,
  citecolor=linkblue,
  urlcolor=linkblue
}

\newcommand{\sectionref}[1]{Sect.~\ref{#1}}
\newcommand{\figref}[1]{Fig.~\ref{#1}}
\newcommand{\tabref}[1]{Table~\ref{#1}}
\newcommand{\appref}[1]{Appendix~\ref{#1}}
\newcommand{\BigO}{\mathcal{O}}
\newcommand{\dd}{\mathop{}\!\mathrm{d}}
\newcommand{\trans}{^{\mathrm T}}

\DeclareMathOperator*{\vecspan}{span}
\DeclareMathOperator*{\atanh}{atanh}

\begin{document}

\title{Natural coordinates for constrained correlation functions:
  Partial autocorrelations and the geometry of positive power spectra}

\author{Thomas Erben\orcidlink{0009-0000-6513-0250}}
\email{terben@astro.uni-bonn.de}
\affiliation{Argelander Institut f\"ur Astronomie, Auf dem H\"ugel
  71, D-53121 Bonn}
\affiliation{Cluster of Excellence “Our Dynamic Universe” (DYNAVERSE)}

\begin{abstract}
  Two-point correlation functions are a standard summary statistic in
  cosmic shear and large-scale-structure analyses. The values they
  may take are, however, constrained: non-negativity of the underlying
  power spectrum restricts any admissible sequence of correlation
  coefficients \((r_1,\ldots,r_N)\) to a bounded convex region,
  described in the one-dimensional case by the recursive interval
  geometry of \citet{SchneiderHartlap2009}. Their formalism introduces
  an affine variable \(x_n\) that maps the admissible interval for
  \(r_n\), given \(r_1,\ldots,r_{n-1}\), to \([-1,+1]\); the algebraic
  status of this variable has so far remained implicit. We identify
  \(x_n\) with the partial autocorrelation coefficient \(\alpha_n\) of
  the associated positive Toeplitz correlation matrix -- a classical
  quantity in time-series analysis whose role in
  constrained-correlation problems has not been recognised in this
  context. The partial autocorrelations thus provide natural
  coordinates on the admissible region, each \(\alpha_n\) varying
  independently in \([-1,+1]\).

  The identification places the standard partial-autocorrelation
  toolbox at direct disposal. The admissible intervals for \(r_n\),
  computed in the original formalism through determinant identities
  and symbolic computer algebra and feasible only at modest order,
  follow now from an \(\BigO(N^2)\) recursion at arbitrary order. The
  inverse hyperbolic tangent applied in the quasi-Gaussian likelihood
  construction of \citet{WilkingSchneider2013} is now identified as
  Fisher's \(z\)-transformation of partial autocorrelations, providing
  a classical statistical interpretation of its empirical
  Gaussianising effect. Numerical experiments illustrate the practical
  use of the natural-coordinate formulation in the tested
  one-dimensional settings. Higher-dimensional isotropic constraints
  lie outside the direct scope of this identification and require
  additional geometric input.
\end{abstract}

\begin{keywords}
    {methods: statistical -- correlation functions --
      Toeplitz matrices -- covariance matrices}
\end{keywords}

\maketitle

\section{Introduction}
\label{sec:introduction}

Two-point correlation functions are among the primary observables
linking cosmological models to sky measurements. In weak gravitational
lensing, the shear correlation functions \(\xi_\pm(\theta)\) measured
across a range of angular separations serve as the immediate data
vector for cosmological parameter inference \citep{Kaiser1992,
  BartelmannSchneider2001, Fu2008, Kilbinger2015}; comparable roles
are played by correlation functions of galaxy positions and by tracer
statistics in other cosmological settings.  Within a Bayesian
framework, converting such a measurement into constraints on model
parameters requires a likelihood, and the default choice throughout
the field is a multivariate Gaussian parametrised by a fiducial mean
vector and a covariance matrix. This convention underlies most cosmic
shear inference pipelines currently in use, including those of major
Stage-III surveys \citep[e.g.][]{Asgari2021, Secco2022, Li2023,
  Wright2025} as well as standard forecast frameworks for upcoming
Stage-IV experiments \citep[e.g.][]{Blanchard2020}.

This default is not without cost. \citet{Hartlap2009} demonstrated
significant non-Gaussianity in the cosmic shear likelihood measured on
ray-tracing simulations, with corresponding effects on the inferred
parameter uncertainties. \citet{SchneiderHartlap2009}, whose formalism
we hereafter denote by SH, traced this failure to a structural fact:
valid correlation-coefficient sequences \((r_1,\ldots,r_N)\) do not
fill the cube \([-1,1]^N\) freely but occupy a bounded convex region
whose boundary is nonlinear and whose interior constraints couple
\(r_n\) recursively to \(r_1,\ldots,r_{n-1}\), taking the form
\(r_{n,\mathrm{l}} \leq r_n \leq r_{n,\mathrm{u}}\) with interval
endpoints determined by the earlier coefficients. Any multivariate
Gaussian likelihood has support on all of \(\mathbb{R}^N\) and
therefore assigns nonzero probability to configurations that lie
outside this admissible region -- configurations that no positive
power spectrum can realise.

\citet{SchneiderHartlap2009} themselves introduced a nonlinear
reparametrisation of the admissible interval that maps each \(r_n\)
onto \([-1,+1]\), and composed it with the inverse hyperbolic tangent
to obtain unbounded coordinates on the real line.
\citet{KeitelSchneider2011} derived the exact univariate and bivariate
probability distributions of correlation coefficients for
one-dimensional Gaussian random fields; the multivariate case is
analytically intractable. \citet{WilkingSchneider2013} combined the
two developments into what they termed the \emph{quasi-Gaussian}
likelihood: a multivariate Gaussian in the transformed coordinates,
transported back to \(\xi\)-space through the univariate marginal
from \citet{KeitelSchneider2011} and the Jacobian of the coordinate
change. Their choice of the inverse hyperbolic tangent was
empirically motivated. \citet{WilkingRoeselerSchneider2015} verified
on halo catalogues from the Millennium Simulation that measured
correlation functions respect the SH bounds and
that the transformed variables exhibit substantially reduced
non-Gaussian signatures.

A complementary harmonic-space approach by \citet{OehlTroster2025} and
\citet{OehlTroster2026} targets the same non-Gaussianity directly,
computing the exact multivariate correlation-function likelihood on
masked spherical Gaussian random fields via the characteristic
function of the underlying quadratic form and a Gaussian-copula
construction.

The idea of parametrising the admissible interval for each \(r_n\) by
a variable in \((-1,+1)\) is not new. In stationary time-series
analysis the same construction appears as the partial autocorrelation
function, introduced by \citet{Levinson1947} and \citet{Durbin1960} in
the context of linear prediction and characterised by
\citet{Ramsey1974} and \citet{BarndorffNielsenSchou1973};
\citet{BrockwellDavis2016} provides a modern textbook account. The
same coefficients arise in the theory of orthogonal polynomials on the
unit circle, where they are known as Verblunsky or Schur reflection
coefficients \citep{Simon2005}. Modern accounts of these ideas for
general correlation matrices and in time-series analysis are given by
\citet{ForresterZhang2020} and \citet{DingZhou2024}, respectively.  To
our knowledge, the connection between this classical framework and the
likelihood modelling of cosmological correlation functions developed
by \citet{SchneiderHartlap2009}, \citet{WilkingSchneider2013}, and
\citet{WilkingRoeselerSchneider2015} has not previously been made
explicit.

Our main result is that the nonlinear reparametrisation onto
\((-1,+1)\) introduced by \citet{SchneiderHartlap2009} coincides
exactly with the classical partial autocorrelation function of the
underlying positive Toeplitz matrix, equivalently with the Schur
(Verblunsky) reflection coefficient. The proof of this identification
turns the recursive constraint problem into a standard
Levinson--Durbin setting.

The scope of this paper is the one-dimensional Toeplitz case in the
normalised coefficients \(r_n = \xi(n\,\Delta x)/\xi(0)\). Two
ingredients needed for a full \(\xi\)-space likelihood -- the
distribution of \(\xi_0\) and the conditioning on it -- remain as in
\citet{WilkingSchneider2013}, and genuinely higher-dimensional
isotropic constraints require additional structure beyond the
one-dimensional Toeplitz property; both points are addressed in the
outlook of \sectionref{sec:conclusions}. The paper is organised as
follows. \sectionref{sec:ccf} sets up the framework and states the
main theorem, whose proof is given in
\sectionref{sec:proof}. \sectionref{sec:consequences} works out the
consequences of the identification,
\sectionref{sec:num_demonstrations} illustrates them numerically, and
\sectionref{sec:conclusions} concludes with an outlook that includes
higher-dimensional directions.

\section{Constrained correlations, Toeplitz matrices, and the Schneider--Hartlap transformation}
\label{sec:ccf}

\subsection{Correlation functions and Toeplitz matrices}
\label{subsec:ccf_setup}

Throughout this paper we consider a statistically homogeneous
one-dimensional real-valued random field \(g(x)\) with vanishing mean
and nonzero variance: \(\mathbb{E}[g(x)] = 0\) and
\(0 < \mathbb{E}[g(x)^{2}]<\infty\). Its correlation function is
\[
  \xi(x) = \langle g(y)\,g(y+x)\rangle,
\]
which depends only on the separation. For a fixed reference separation
\(\Delta x > 0\) we form the normalised correlation coefficients
\[
  r_n := \frac{\xi(n\,\Delta x)}{\xi(0)}, \qquad r_0 = 1,
\]
where \(n=1,2,\ldots\) enumerates integer multiples of the base
separation. In an astronomical context, one may think of the \(r_n\)
as the shear correlation coefficients measured at multiples of a fixed
angular scale, or as the correlation coefficients of any other
homogeneous tracer sampled on an equidistant grid.

It is convenient to introduce a Hilbert-space realisation of these
correlations from the outset and we define the unit-normalised field
values
\[
  e_j := \frac{g(j\,\Delta x)}{\sqrt{\xi(0)}}, \qquad j\in\mathbb{N}_{0},
\]
regarded as elements of the Hilbert space of centred,
square-integrable random variables, equipped with the inner
product \(\langle X,Y\rangle = \mathbb{E}[XY]\). Then \(\|e_j\|=1\),
and the inner products reproduce the correlation coefficients,
\[
  \langle e_i,e_j\rangle = r_{|i-j|}.
\]
The Gram matrix of any finite collection
\(e_0,e_1,\ldots,e_{n-1}\) is the finite Toeplitz correlation matrix
\[
  A_n = \begin{pmatrix}
  1      & r_1    & r_2    & \cdots & r_{n-1} \\
  r_1    & 1      & r_1    & \cdots & r_{n-2} \\
  r_2    & r_1    & 1      & \cdots & r_{n-3} \\
  \vdots & \vdots & \vdots & \ddots & \vdots  \\
  r_{n-1} & r_{n-2} & r_{n-3} & \cdots & 1
  \end{pmatrix} = \bigl(r_{|i-j|}\bigr)_{i,j=0}^{n-1},
\]
so that \(A_n\) contains the coefficients \(r_0,r_1,\ldots,r_{n-1}\).
Throughout the paper we write \(A\succ 0\) for a positive definite
matrix and \(A\succeq 0\) for a positive semidefinite matrix, in the
sense of the Löwner order on symmetric matrices. Because \(A_n\) is a
Gram matrix, \(A_n\succeq 0\).

The positive semidefiniteness of the finite Toeplitz matrices \(A_n\)
is the finite-dimensional expression of the nonnegativity of the
underlying power spectrum \citep{SchneiderHartlap2009}. Throughout the
following sections and appendices we fix a positive integer \(N\) and
call a correlation-coefficient vector \((r_1, \ldots, r_N)\)
\emph{admissible} if its associated Toeplitz matrices \(A_n\) are
positive semidefinite for all \(1 \leq n \leq N+1\). The admissible
vectors form a convex subset of \([-1,1]^N\) whose recursive geometry
is the subject of this paper.

\subsection{The Schneider--Hartlap admissible interval}
\label{subsec:ccf_sh}

\citet{SchneiderHartlap2009} showed that the nonnegativity constraints
\(A_n \succeq 0\) do not merely restrict the correlation coefficients
\(r_n\) to a fixed range but rather couple them nontrivially.

The starting point is the recursive nature of the admissibility
constraints. Suppose the coefficients \(r_1,\ldots,r_{n-1}\) have
already been chosen such that the corresponding Toeplitz matrix
\(A_n\) is positive semidefinite. The next coefficient \(r_n\) is
admissible precisely if the enlarged matrix \(A_{n+1}\), which
additionally contains \(r_n\) along its outermost diagonal, remains
positive semidefinite. As shown by \citet{SchneiderHartlap2009}, this
condition restricts \(r_n\) to a closed interval
\[
  r_{n,\mathrm{l}} \leq r_n \leq r_{n,\mathrm{u}},
\]
whose bounds are functions of the previously specified
\(r_1,\ldots,r_{n-1}\) alone. Concrete expressions for these bounds,
derived via covariance-matrix determinants and via the Cauchy--Schwarz
inequality, are given explicitly for \(n\leq 4\) in
\citet{SchneiderHartlap2009}; for larger \(n\), the bounds are of
increasing algebraic complexity and symbolic calculations were done
with computer algebra up to \(n=16\). Two identities relating the
interval widths \(\Delta_n = r_{n,\mathrm{u}} - r_{n,\mathrm{l}} \) to
the associated Toeplitz matrices (their Eqs.~24, 25, and 38) were
established there by direct verification. A rigorous proof of these
identities, without reference to computer algebra, was given in
\citet{Erben2026}.

To turn this recursive structure into a set of coordinates on the
admissible region, \citet{SchneiderHartlap2009} introduced the
normalised interval variable, whose endpoints \(r_{n,\mathrm{l}}\) and
\(r_{n,\mathrm{u}}\) depend only on the previously specified
\(r_1,\ldots, r_{n-1}\),
\[
  x_n
  =
  \frac{2\,r_n - r_{n,\mathrm{u}} - r_{n,\mathrm{l}}}
       {r_{n,\mathrm{u}} - r_{n,\mathrm{l}}},
\]
which maps the admissible interval for \(r_n\) onto \([-1,1]\), and
subsequently applied the Fisher-type transformation
\[
  y_n = \atanh(x_n)
\]
to obtain unbounded coordinates \(y_n\) on the real line. The
coordinates \(y_n\) were shown by
\citet{WilkingRoeselerSchneider2015}, using one-dimensional Gaussian
random field simulations and a three-dimensional test on the
Millennium Simulation halo catalogue, to render the joint distribution
of correlation coefficients substantially more Gaussian than in the
original \(r_n\) representation. This quasi-Gaussian behaviour
underlies the practical appeal of the SH coordinates
for likelihood modelling of constrained correlation functions.

The practical usefulness of the original SH parametrisation is limited
by the algebraic complexity of the bounds \(r_{n,\mathrm{l}}\) and
\(r_{n,\mathrm{u}}\), which grow rapidly with \(n\) in the original
\(r\)-coordinates. Independently, the SH variable \(x_n\) at first
appears as a handy but arbitrary rescaling of the admissible
interval. Surprisingly, both aspects look different from the
perspective of classical time-series analysis. There, positive
Toeplitz matrices admit a canonical parametrisation by their partial
autocorrelation coefficients through the Levinson--Durbin recursion
\citep{Levinson1947,Durbin1960}, which reduces the algebraic
complexity to an \(\BigO(N^2)\) forward pass and equips \(x_n\) with
an intrinsic statistical meaning. We make this identification precise
in \sectionref{subsec:ccf_main_theorem} and work out its consequences
in \sectionref{sec:consequences}.

\subsection{Partial autocorrelations from linear prediction}
\label{subsec:ccf_pacf}

A cosmic shear correlation function measured at angular
separations \(\theta, 2\theta, 3\theta, \ldots\) contains
redundancy: the correlation \(r_3\) at separation \(3\theta\) is not
statistically independent of \(r_1\) and \(r_2\). Part of what
\(r_3\) captures is the direct correlation between field points
separated by \(3\theta\), but another part is transmitted through
the intermediate scales at \(\theta\) and \(2\theta\). Isolating the
direct contribution requires removing this transmitted part
explicitly.

The natural way to do so is by linear prediction. Given the previously
specified coefficients \(r_1, \ldots, r_{n-1}\), let \(p_n\) denote
the endpoint correlation linearly predicted by them in the
least-squares sense, and let \(\sigma_n^2\) denote the associated
residual variance. Both quantities depend only on \(r_1, \ldots,
r_{n-1}\) and can be computed explicitly from the Toeplitz sub-block
\(B = (r_{|i-j|})_{i,j=1}^{n-1}\). The partial autocorrelation
coefficient \(\alpha_n\) at lag \(n\) is defined as the normalised
residual,
\[
  \alpha_n = \frac{r_n - p_n}{\sigma_n^2}.
\]
It measures the part of the endpoint correlation \(r_n\) that cannot
be predicted linearly from the intermediate correlations.

Geometrically, the same object has a compact interpretation in terms
of the vectors \(e_j\) introduced in \sectionref{subsec:ccf_setup}. In
this Hilbert space the best linear predictor coincides with the
orthogonal projection onto the subspace
\(V=\vecspan(e_1,\ldots,e_{n-1})\) spanned by the intermediate
vectors, and the residual variance \(\sigma_n^2\) coincides with the
squared norm of the residual of \(e_0\) after projection. Both
endpoint vectors \(e_0\) and \(e_n\) decompose into their projections
onto \(V\), which are determined entirely by \(r_1, \ldots, r_{n-1}\),
and orthogonal residuals \(\widetilde e_0\) and \(\widetilde e_n\),
which carry the entire remaining freedom. These two residuals have the
same norm \(\sigma_n\), a consequence of the Toeplitz structure of
\(A_n\) established in the proof within \sectionref{sec:proof}.  In
this picture \(\alpha_n\) is the correlation coefficient between the
two endpoint residuals,
\[
  \alpha_n
  =
  \frac{\langle \widetilde e_0, \widetilde e_n\rangle}
       {\|\widetilde e_0\|\,\|\widetilde e_n\|}
  = \cos\bigl(\sphericalangle(\widetilde e_0,\widetilde e_n)\bigr),
\]
that is, the cosine of the angle between them once the intermediate
subspace has been projected out. Equivalently, \(\alpha_n\) is the
off-diagonal entry of the normalised \(2\times 2\) Gram matrix of the
two endpoint residuals. The admissibility condition on \(r_n\)
translates into the positive semidefiniteness of this small matrix,
which in turn is equivalent to \(|\alpha_n|\leq 1\). This geometric
formulation captures the essential content of the main theorem and
will carry the proof in \sectionref{sec:proof}.

Two low-order cases illustrate the construction and confirm its
consistency with the SH interval. For \(n=1\) there are
no intermediate coefficients, so \(p_1 = 0\), \(\sigma_1^2 = 1\), and
\(\alpha_1 = r_1\). For \(n=2\) the Toeplitz sub-block reduces to the
scalar \(B = 1\), the predicted correlation is \(p_2 = r_1^2\), the
residual variance is \(\sigma_2^2 = 1 - r_1^2\), and hence
\[
  \alpha_2 = \frac{r_2 - r_1^2}{1 - r_1^2}.
\]
The admissibility interval for \(r_2\) derived by Schneider and
Hartlap, \(-1 + 2r_1^2 \leq r_2 \leq 1\), maps under this
transformation to \(-1 \leq \alpha_2 \leq 1\). The interval condition
on \(r_2\) becomes a bound on a single variable in a fixed range. This
is already the main theorem of the present paper in its simplest
nontrivial instance.

The coefficient \(\alpha_{n}\) is the classical partial
autocorrelation coefficient of stationary time-series analysis,
sometimes denoted \(\kappa_{n}\) and also sign conventions might vary
in the literature. We use \(\alpha_{n}\) throughout, with the sign
convention fixed by the definition above.

\subsection{Statement of the main theorem}
\label{subsec:ccf_main_theorem}

We are now in a position to state the main result of this paper.

\paragraph{Main Theorem.}
Let \((r_1, \ldots, r_N)\) be an admissible correlation vector with
associated Toeplitz matrices \(A_n \succeq 0\), and let \(x_n\) denote
the SH coordinate at lag \(n\). Then for every \(n\) at which the
previous Toeplitz matrix \(A_n\) is positive definite,
\[
  \boxed{x_n = \alpha_n},
\]
where \(\alpha_n\) is the \(n\)-th partial autocorrelation coefficient
of the sequence.

At the boundary of the admissible region, where \(A_n\) first becomes
singular, the residual variance \(\sigma_n^2\) vanishes and neither
\(\alpha_n\) nor \(x_n\) is defined; this case is treated in
\sectionref{subsec:proof_degen}. Note explicitly that, at this stage,
no statement is made on the existence of the underlying correlation
sequence \(r_n\) past this point. This will be developed in
\sectionref{subsec:cons_boundary}. \tabref{tab:dictionary} summarises the
correspondence between the SH notation and the
classical partial-autocorrelation notation used throughout the paper.

\begin{table*}[t]
  \caption{Correspondence between the SH coordinates and the classical
    partial-autocorrelation coordinates used in this paper.  The
    identifications hold in the non-degenerate case (\(A_n \succ 0\));
    at the boundary of the admissible region, neither \(\alpha_n\) nor
    \(x_n\) is defined (see \sectionref{subsec:proof_degen}). The
    symbol \(\kappa_n\) is also common in the time-series literature.}
  \label{tab:dictionary}
  \centering
  \begin{tabular}{lll}
    \hline\hline
    Schneider--Hartlap & Present paper / classical & Meaning \\
    \hline
    \(r_n\)                    & \(r_n\)                & correlation at lag \(n\) \\
    \(r_{n,\mathrm{l}},r_{n,\mathrm{u}}\) & \(p_n \pm \sigma_n^2\) & admissible interval endpoints \\
    \(x_n\)                    & \(\alpha_n = \kappa_n\) & partial autocorrelation \\
    \(y_n = \atanh(x_n)\) & \(\atanh(\alpha_n)\) & Fisher \(z\)-coordinate \\
    \(\Delta_n = r_{n,\mathrm{u}} - r_{n,\mathrm{l}}\) & \(2\sigma_n^2\) & interval width \\
    --                         & \(p_n\)                & predicted correlation \\
    --                         & \(\sigma_n^2\)         & residual variance \\
    \hline
  \end{tabular}
\end{table*}

\section{Proof of the main theorem}
\label{sec:proof}

We prove the identification \(x_n=\alpha_n\) stated in
\sectionref{subsec:ccf_main_theorem}. The argument splits into two
parts: the non-degenerate case and the boundary case, where a
Toeplitz matrix becomes singular for the first time.

\subsection{The non-degenerate case}
\label{subsec:proof_non_degen}

We assume throughout this subsection that \(A_n\succ0\).

\paragraph{The case \(n=1\).}
For \(n=1\), the enlarged Toeplitz matrix is
\[
A_2=
  \begin{pmatrix}
    1 & r_1\\
    r_1 & 1
  \end{pmatrix}.
\]
The condition \(A_2\succeq0\) is equivalent to
\[
-1\leq r_1\leq1.
\]
Thus \(r_{1,\mathrm{l}}=-1\), \(r_{1,\mathrm{u}}=1\), and
\[
x_1=r_1.
\]
Since there are no intermediate vectors, the first partial
autocorrelation is also \(\alpha_1=r_1\). Hence \(x_1=\alpha_1\). We
treated the \(n=1\) case separately to avoid introducing an empty Gram
matrix in \eqref{eq:proof_B} below.

\paragraph{Setup for \(n\geq2\).}
We now assume \(n\geq2\). Since \(A_n\succ0\), it admits a Gram
realisation by linearly independent vectors and we therefore choose
\(e_0,\ldots,e_{n-1}\) with Gram matrix \(A_n\). Any two such
realisations are isometric; we fix this choice and construct all
further vectors as extensions of it. Let
\[
V=\vecspan(e_1,\ldots,e_{n-1})
\]
and write \(E\) for the linear map whose columns are the intermediate
vectors,
\[
E=
  \begin{pmatrix}
  | &        & |\\
  e_1 & \cdots & e_{n-1}\\
  | &        & |
  \end{pmatrix}.
\]

The Gram matrix of the intermediate vectors is
\begin{equation}
  B:=E\trans E=(r_{|i-j|})_{i,j=1}^{n-1}.
  \label{eq:proof_B}
\end{equation}
Since \(A_n\succ 0\), also \(B\succ 0\). Hence the orthogonal projection
onto \(V\) is
\begin{equation}
  P_V=E(E\trans E)^{-1}E\trans =EB^{-1}E\trans.
  \label{eq:proof_projection}
\end{equation}
We further define
\begin{equation}
  u:=E\trans e_0=(r_1,\ldots,r_{n-1})\trans,
  \qquad
  v:=(r_{n-1},\ldots,r_1)\trans.
  \label{eq:proof_uv}
\end{equation}
If \(r_n\) is admissible and \(e_n\) is an endpoint vector completing
the Gram matrix \(A_{n+1}(r_n)\), then \(E\trans e_n=v\).

\paragraph{Necessity of the interval.}
Let \(r_n\) be admissible, and let \(e_n\) be such that
\(e_0,\ldots,e_n\) have Gram matrix \(A_{n+1}(r_n)\). By
\eqref{eq:proof_projection},
\[
P_Ve_0=EB^{-1}u,
  \qquad
  P_Ve_n=EB^{-1}v.
\]
Thus the residuals after projecting out the intermediate subspace are
\begin{equation}
  \widetilde e_0=e_0-P_Ve_0=e_0-EB^{-1}u,
  \quad
  \widetilde e_n=e_n-EB^{-1}v.
  \label{eq:proof_residuals}
\end{equation}
We compute their inner product:
\[
\begin{aligned}
\langle \widetilde e_0,\widetilde e_n\rangle
&= \langle e_0-EB^{-1}u,\,
          e_n-EB^{-1}v\rangle \\
&= \langle e_0,e_n\rangle
   - \langle e_0,EB^{-1}v\rangle \\
&\quad
   - \langle EB^{-1}u,e_n\rangle
   + \langle EB^{-1}u,EB^{-1}v\rangle \\
&= r_n-u\trans B^{-1}v-u\trans B^{-1}v
   +u\trans B^{-1}BB^{-1}v \\
&= r_n-u\trans B^{-1}v .
\end{aligned}
\]
Thus, with
\begin{equation}
  p_n:=u\trans B^{-1}v,
  \label{eq:proof_pn}
\end{equation}
we have
\begin{equation}
  \langle \widetilde e_0,\widetilde e_n\rangle=r_n-p_n.
  \label{eq:proof_residual_inner_product}
\end{equation}
Similarly,
\[
\begin{aligned}
  \|\widetilde e_0\|^2
  &= \langle e_0-EB^{-1}u,\,
             e_0-EB^{-1}u\rangle\\
  &= 1 - u\trans B^{-1}u
       - u\trans B^{-1}u
       + u\trans B^{-1}BB^{-1}u\\
  &= 1 - u\trans B^{-1}u.
\end{aligned}
\]
We therefore define the residual variance by
\begin{equation}
  \sigma_n^2:=\|\widetilde e_0\|^2=1 - u\trans B^{-1}u.
  \label{eq:proof_sigma}
\end{equation}
Geometrically, \(\sigma_n^2\) measures how much of \(e_0\) lies outside
the subspace spanned by the intermediate vectors.
Since \(A_n\succ0\), the vectors \(e_0,\ldots,e_{n-1}\) are linearly
independent. Hence \(e_0\notin V\), and therefore
\[
\sigma_n^2>0.
\]
Next, we show that the residual of the other endpoint has the
same norm. Let \(J\) be the \((n-1)\times(n-1)\) reversal matrix,
\begin{equation}
  J=(\delta_{i,n-j})_{1\leq i,j\leq n-1}
  =
  \begin{pmatrix}
  0      & \cdots & 0      & 1 \\
  \vdots & \reflectbox{\(\ddots\)} & \vdots & \vdots \\
  0      & 1      & \cdots & 0 \\
  1      & 0      & \cdots & 0
  \end{pmatrix}.
  \label{eq:proof_reversal_matrix}
\end{equation}
Then \(J\trans =J=J^{-1}\) and \(v=Ju\). Since \(B\) is Toeplitz,
\begin{equation}
  \begin{aligned}
    (JBJ)_{ij}
    &= \sum_{k,l=1}^{n-1} J_{ik}B_{kl}J_{lj} \\
    &= \sum_{k,l=1}^{n-1}
       \delta_{i,n-k}B_{kl}\delta_{l,n-j} \\
    &= B_{n-i,n-j}
     = r_{|(n-i)-(n-j)|}
     = r_{|i-j|}
     = B_{ij}.
  \end{aligned}
  \label{eq:proof_JBJ}
\end{equation}
Thus \(JBJ=B\). Since \(J^{-1}=J\), inversion gives
\[
  JB^{-1}J=B^{-1}.
\]
Consequently,
\[
  v\trans B^{-1}v=u\trans J\trans B^{-1}Ju=u\trans B^{-1}u.
\]
The second endpoint residual therefore has norm
\[
  \|\widetilde e_n\|^2 = 1 - v\trans B^{-1}v = 1 - u\trans B^{-1}u = \sigma_n^2.
\]
The Gram matrix of the two endpoint residuals is thus
\begin{equation}
G_n=
  \begin{pmatrix}
    \langle \widetilde e_0,\widetilde e_0\rangle &
    \langle \widetilde e_0,\widetilde e_n\rangle\\
    \langle \widetilde e_n,\widetilde e_0\rangle &
    \langle \widetilde e_n,\widetilde e_n\rangle
  \end{pmatrix} =
  \begin{pmatrix}
    \sigma_n^2 & r_n-p_n\\
    r_n-p_n & \sigma_n^2
  \end{pmatrix}.
  \label{eq:proof_Gn}
\end{equation}
For the admissible \(r_n\) under consideration, \(G_n\) is an actual
Gram matrix, and hence \(G_n\succeq0\). Its eigenvalues are
\[
  \lambda_\pm=\sigma_n^2\pm(r_n-p_n).
\]
The condition \(G_n\succeq0\) therefore implies
\begin{equation}
  p_n-\sigma_n^2\leq r_n\leq p_n+\sigma_n^2.
  \label{eq:proof_interval_necessary}
\end{equation}
Thus every admissible value of \(r_n\) lies in the closed interval
shown in \eqref{eq:proof_interval_necessary}.

\paragraph{Sufficiency of the interval.}
We now prove the converse: no further constraints are hidden and
\emph{each} \(r_n\) defined by \eqref{eq:proof_interval_necessary} is
indeed admissible.

The map \(r_n \mapsto A_{n+1}(r_n)\) is affine, so the set of
admissible values of \(r_n\) is convex: if \(r_n^{(0)}\) and
\(r_n^{(1)}\) are both admissible, then for every
\(\lambda \in [0,1]\),
\begin{equation}
\begin{aligned}
  A_{n+1}\bigl(\lambda r_n^{(0)} + (1 - \lambda) r_n^{(1)}\bigr)
  &={} \lambda\, A_{n+1}(r_n^{(0)}) \\
  &\quad {}+ (1 - \lambda)\, A_{n+1}(r_n^{(1)})
  \succeq 0.
\end{aligned}
\label{eq:proof_convexity}
\end{equation}
It therefore suffices to realise the two endpoints
\(r_n^{\pm} := p_n \pm \sigma_n^2\) of the interval
\eqref{eq:proof_interval_necessary}; every intermediate value then
follows by convexity.

For the endpoints, consider the two candidate vectors
\begin{equation}
  e_n^{\pm} := EB^{-1}v \pm \widetilde e_0.
  \label{eq:proof_endpoint_vectors}
\end{equation}
Both lie in \(\vecspan(e_0, e_1, \ldots, e_{n-1})\) and require no
enlargement of the Hilbert space. Using the relations already derived,
namely
\[
  E\trans  \widetilde e_0 = 0,\;
  E\trans  E = B,\;
  E\trans  e_0 = u,\;
  v\trans  B^{-1} v = u\trans  B^{-1} u,
\]
a short calculation gives
\[
  E\trans  e_n^{\pm} = v,
  \quad
  \|e_n^{\pm}\|^2 = 1,
  \quad
  \langle e_0, e_n^{\pm}\rangle = p_n \pm \sigma_n^2.
\]
Thus \(e_0, \ldots, e_{n-1}, e_n^{\pm}\) realise the Toeplitz matrices
\(A_{n+1}(r_n^{\pm})\), and both endpoints of the interval are
admissible.

Combining necessity with convexity and the endpoint construction, the
admissible interval is exactly
\begin{equation}
  r_{n,\mathrm{l}}=p_n-\sigma_n^2,
  \qquad
  r_{n,\mathrm{u}}=p_n+\sigma_n^2.
  \label{eq:proof_interval_endpoints}
\end{equation}

\paragraph{Identification with the partial autocorrelation.}
Substituting \eqref{eq:proof_interval_endpoints} into the
SH variable gives
\[
\begin{aligned}
  x_n
  &=
  \frac{2r_n-r_{n,\mathrm{u}}-r_{n,\mathrm{l}}}
       {r_{n,\mathrm{u}}-r_{n,\mathrm{l}}}\\
  &=
  \frac{r_n-p_n}{\sigma_n^2}.
\end{aligned}
\]
On the other hand, by definition of the partial autocorrelation and by
\eqref{eq:proof_residual_inner_product} and \eqref{eq:proof_sigma},
\[
  \alpha_n
    =
    \frac{\langle\widetilde e_0,\widetilde e_n\rangle}
         {\|\widetilde e_0\|\,\|\widetilde e_n\|}
    =
    \frac{r_n-p_n}{\sigma_n^2}.
\]
Therefore
\begin{equation}
  x_n=\alpha_n.
  \label{eq:proof_main_identity}
\end{equation}

Equivalently, normalising the residual Gram matrix
\eqref{eq:proof_Gn} by \(\sigma_n^2\) gives
\[
  \frac{1}{\sigma_n^2}G_n
    =
    \begin{pmatrix}
      1 & x_n\\
      x_n & 1
    \end{pmatrix}
    =
    \begin{pmatrix}
      1 & \alpha_n\\
      \alpha_n & 1
    \end{pmatrix}.
\]

The SH coordinate is therefore not a different
nonlinear parametrisation; it is precisely the partial autocorrelation
of the positive Toeplitz matrix, viewed geometrically as the residual
correlation once the intermediate subspace has been projected out.

\subsection{The degenerate case}
\label{subsec:proof_degen}

The preceding argument assumes \(A_n\succ 0\). We now describe what
happens at the positive semidefinite boundary.

First, we must fix the boundary-index convention used
throughout the remainder of the paper. Following
\citet{SchneiderHartlap2009}, we anchor the boundary index on the
Toeplitz side: \emph{\(m\) denotes the index of the first singular
Toeplitz matrix \(A_m\)}. Since \(A_m\) contains the coefficients up to
\(r_{m-1}\), the coordinate \(x_m\) is the first no longer defined by
the non-degenerate construction of
\sectionref{subsec:proof_non_degen}. This termination of the SH
\(x\)-chart at \(A_m\) was established in \citet{Erben2026}.
We note that the PACF literature more commonly anchors the boundary on the
coefficient side, using the last well-defined \(\alpha_{m-1}\); since
\(\alpha_n\) parametrises the passage from \(A_n\) to \(A_{n+1}\),
this labels the \emph{same} transition with an index shifted by one.

The sequence-side consequences for \(r_{n}\) -- collapse of the subsequent SH
intervals to single points and deterministic continuation of the
correlation sequence -- are discussed in
\sectionref{subsec:cons_boundary}. Here we establish the corresponding
PACF-side statement by showing \(\sigma_m^{2} = 0\).

Since \(A_m\succeq 0\), it can be realised as a Gram matrix, although
the vectors are now linearly dependent. Since \(A_m\) is singular,
there exists a nonzero vector
\[
  w=(w_0,\ldots,w_{m-1})\trans
\]
such that \(A_mw=0\). Moreover \(w_0\neq0\). Indeed, if \(w_0=0\),
then \((w_1,\ldots,w_{m-1})\trans \) would be a null vector of the
lower-right principal block of \(A_m\). This block is \(A_{m-1}\),
contradicting \(A_{m-1}\succ 0\).

Using the Gram representation, we obtain
\[
  0
    =
    w\trans A_mw
    =
    \left\|\sum_{j=0}^{m-1}w_j e_j\right\|^2 \Rightarrow
    \sum_{j=0}^{m-1}w_j e_j=0.
\]
Since \(w_0\neq 0\), this implies
\[
  e_0
    =
    -\frac{1}{w_0}\sum_{j=1}^{m-1}w_j e_j
    \in
    \vecspan(e_1,\ldots,e_{m-1}).
\]
At step \(m\), the intermediate subspace is
\[
  V_m=\vecspan(e_1,\ldots,e_{m-1}).
\]
Thus \(e_0\in V_m\). Therefore its residual with respect to this
subspace vanishes,
\[
  \widetilde e_0^{(m)}
    =
    e_0-P_{V_m}e_0
    =
    0,
\]
and hence
\[
  \sigma_m^2 = \|\widetilde e_0^{(m)}\|^2 = 0.
\]
For any admissible continuation beyond this point, the intermediate
subspace only becomes larger and still contains \(e_0\). Therefore
\[
  \sigma_{m+s}^2=0,\qquad s\geq 0.
\]

With \(\sigma_m^{2} = 0\), the definition
\(\alpha_m = (r_m - p_m)/\sigma_m^{2}\) of
\sectionref{subsec:ccf_pacf} fails: the normalised partial
autocorrelation of order \(m\) is no longer defined, and the last
well-defined PACF coefficient is \(\alpha_{m-1}\). The PACF chart
therefore reaches its boundary at the same Toeplitz-matrix index
\(m\) as the SH \(x\)-chart.

\section{Consequences}
\label{sec:consequences}

Having established the equivalence between the SH
coordinates and the classical partial autocorrelations, we now
reformulate the first in the language of the latter.  Rather than
surveying the theory of partial autocorrelation functions,
we focus on the aspects directly relevant to the SH
framework and to the likelihood modelling of one-dimensional
constrained correlation functions: the coordinate bijection, the
efficient computation of bounds via the classical Levinson--Durbin
recursion, the Fisher interpretation of the quasi-Gaussian variables,
the Jacobian for density transport, and the induced volume and
boundary structure of the admissible region.

\subsection{Bijective coordinates on the admissible region}
\label{subsec:cons_bijection}

By the main theorem, the SH coordinate \(x_n\) coincides with the
classical partial autocorrelation \(\alpha_n\) on the interior of the
admissible region. This coincidence takes the form of a bijective
coordinate transformation
\[
  \alpha_n = \frac{r_n - p_n}{\sigma_n^2},
  \qquad
  r_n = p_n + \alpha_n\,\sigma_n^2,
\]
and, since \(p_n\) and \(\sigma_n^2\) depend only on
\(r_1,\ldots,r_{n-1}\) (cf.\ \sectionref{subsec:ccf_pacf}), the
transformation is triangular in both directions. Every interior
admissible sequence \((r_1,\ldots,r_N)\), with \(A_{N+1}\succ 0\) and
hence \(\sigma_n^2>0\) at every lag, therefore determines a unique
\((\alpha_1,\ldots,\alpha_N)\in(-1,1)^N\), and every point of the open
hypercube generates a unique admissible correlation sequence by the
forward recursion. The boundary case, where some \(\sigma_n^2\)
vanishes, is the degenerate configuration of
\sectionref{subsec:proof_degen} and is revisited in
\sectionref{subsec:cons_boundary}.

In these coordinates the recursively coupled strict inequalities
\(r_{n,\mathrm{l}} < r_n < r_{n,\mathrm{u}}\) become the
independent one-dimensional bounds \(-1<\alpha_n<1\); the complexity
of the admissible region is absorbed into the coordinate
transformation rather than removed. This parametrisation is
admissibility-preserving by construction: \(N\) numbers drawn
independently from \((-1,1)\) and passed through the forward
recursion yield an admissible correlation function every time, in
direct contrast to the rejection-based construction of
\citet{SchneiderHartlap2009}, whose surviving fraction falls off
rapidly with \(N\). The partial autocorrelations therefore provide
natural coordinates for MCMC or prior construction on the space of
admissible correlation functions.

\subsection{The Levinson--Durbin recursion}
\label{subsec:cons_bounds}

The practical computation of the admissible bounds, and more
generally of the coordinate transformation \(r \leftrightarrow \alpha\)
itself, is supplied by the classical Levinson--Durbin recursion
\citep{Levinson1947,Durbin1960}, viewed here from the perspective of
constrained correlation functions. In \(r\)-coordinates the bounds
become symbolically intractable at high order, as documented by
\citet{SchneiderHartlap2009}; in the natural coordinates of
\sectionref{subsec:cons_bijection} both the bounds and the
coordinate transformation collapse into a single forward pass.

To state the recursion, we introduce the coefficients
\(\phi_j^{(n)}\) of the best linear predictor of \(e_0\) from the
intermediate vectors,
\begin{equation}
  P_{V_n} e_0 = \sum_{j=1}^{n-1} \phi_j^{(n)}\, e_j,
  \label{eq:ld_phi_def}
\end{equation}
so that the predicted correlation of
\sectionref{subsec:ccf_pacf} takes the scalar-product form
\(p_n = \sum_j \phi_j^{(n)}\, r_{n-j}\), obtained from
\(p_n = \langle P_{V_n} e_0, e_n\rangle\) together with
\(\langle e_j, e_n\rangle = r_{n-j}\). The Levinson--Durbin
recursion advances all quantities in a single forward step from
order \(n\) to order \(n+1\):
\begin{subequations}
  \label{eq:ld_recursion}
  \begin{align}
    p_n
    &= \sum_{j=1}^{n-1} \phi_j^{(n)}\, r_{n-j},
    \label{eq:ld_p}\\[3pt]
    \alpha_n
    &= \frac{r_n - p_n}{\sigma_n^{2}},
    \label{eq:ld_alpha}\\[3pt]
    \phi_j^{(n+1)}
    &=
    \begin{cases}
      \phi_j^{(n)} - \alpha_n\, \phi_{n-j}^{(n)},
      & j = 1,\ldots,n-1,\\[2pt]
      \alpha_n,
      & j = n,
    \end{cases}
    \label{eq:ld_phi}\\[3pt]
    \sigma_{n+1}^{2}
    &= \sigma_n^{2}\,\bigl(1 - \alpha_n^{2}\bigr).
    \label{eq:ld_sigma}
  \end{align}
\end{subequations}

The recursion is initialised at \(n = 1\) by \(\sigma_1^{2} = 1\),
\(p_1 = 0\), and \(\alpha_1 = r_1\). Its last line,
\eqref{eq:ld_sigma}, is the residual-variance update and carries the
Jacobian, volume, and boundary analyses of the following
subsections. A derivation of the four update rules from the geometry
of \sectionref{subsec:proof_non_degen} is given in
\appref{app:ld_derivation}; further consequences of the
Levinson--Durbin recursion are collected in \appref{app:corollaries}
and used in the following subsections.

The same forward pass simultaneously produces the admissible
SH bounds
\begin{equation}
  r_{n,\mathrm{l}} = p_n - \sigma_n^{2},
  \qquad
  r_{n,\mathrm{u}} = p_n + \sigma_n^{2},
  \label{eq:ld_bounds}
\end{equation}
and the partial autocorrelations \(\alpha_n\), which parametrise the
bijection of \sectionref{subsec:cons_bijection}. Each step involves
only \(\BigO(n)\) arithmetic operations, so all \(N\) pairs of bounds
and PACF coefficients are obtained in a total of \(\BigO(N^2)\)
operations, without symbolic manipulation and without matrix
inversion. The quadratic operation count makes lag numbers around \(N
\approx 50\) computationally realistic in the one-dimensional
setting. The mutual consistency of the two evaluation directions of
the recursion under finite-precision arithmetic is verified in
\sectionref{subsec:num_verification}.

Finally note that the inverse map \(\alpha \to r\) is carried by the
same recursion, with \eqref{eq:ld_alpha} read as \(r_n = p_n +
\alpha_n\,\sigma_n^{2}\) so that \(\alpha_n\) becomes the input and
\(r_n\) the output at step~\(n\), at identical \(\BigO(N^2)\) cost.

\subsection{Fisher coordinates and the mechanism of quasi-Gaussianisation}
\label{subsec:cons_fisher}
The final component of the SH coordinate system maps the admissible
interval to the unbounded real line,
\begin{equation}
  y_n = \atanh(x_n) = \atanh(\alpha_n),
  \label{eq:fisher_z}
\end{equation}
which is Fisher's \(z\)-transformation applied to the \(n\)-th partial
autocorrelation \citep{Fisher1915}. The functional coincidence was
already noted by \citet{SchneiderHartlap2009} as a possible
explanation for the empirical quasi-Gaussianisation later documented
by \citet{WilkingSchneider2013} and
\citet{WilkingRoeselerSchneider2015}; it is now turned from a
coincidence into a structural statement. \(\alpha_n\) is the
correlation coefficient between the two endpoint residuals after the
intermediate field values have been projected out, so the Fisher map
is applied to an intrinsic residual correlation selected by the
Toeplitz geometry, not to an arbitrary interval rescaling.

Read in this light, the two-stage transformation \(r \to \alpha \to
y\) removes the main coordinate-level obstruction to a Gaussian
description by sending the recursively coupled SH bounds to
infinity. The qualifier \emph{quasi-Gaussian} nevertheless remains
essential. The present setting is not identical to Fisher's original
one: the \(\alpha_n\) are not independent sample correlations of a
bivariate normal population but partial autocorrelations of a
Toeplitz-positive correlation sequence, so intrinsic higher-order
structure of the underlying field distribution can remain in
\(y\)-space even after the admissibility constraints have been
absorbed into the coordinate system. The role of the Fisher
coordinates is thus to remove the bounded-domain obstruction to a
Gaussian description of \(p(y)\) by sending the recursively coupled SH
bounds to infinity -- a necessary but not sufficient condition, whose
empirical adequacy in the follow-up literature is a separate
statement.  The closure test of
\sectionref{subsec:num_gaussianisation} examines how much of the
residual non-Gaussianity survives the coordinate change in the
one-dimensional Gaussian-field setting.

\subsection{A product-form Jacobian for likelihood transport}
\label{subsec:cons_jacobian}

Bayesian inference on constrained correlation functions rests on the
change of variables between the correlation coefficients
\(r = (r_1,\ldots,r_N)\) and the unbounded coordinates
\(y = (y_1,\ldots,y_N)\) of \citet{SchneiderHartlap2009}. In the
notation of \citet[Eq.~20]{WilkingSchneider2013},
\[
  p_r(r_1,\ldots,r_N)
  =
  p_y(y_1,\ldots,y_N)\,
  \left|\det\!\left(\frac{\partial y}{\partial r}\right)\right|,
\]
so a Gaussian model in \(y\)-space induces a non-Gaussian likelihood
on the admissible region in \(r\)-space through the Jacobian factor.
\citet[Eq.~21]{WilkingSchneider2013} gave this Jacobian in triangular
form; in natural coordinates, the transformation \(r \to y\) factors
canonically into two stages,
\[
  r \longrightarrow \alpha \longrightarrow y,
  \qquad
  y_n = \atanh(\alpha_n),
\]
each of which admits a product-form Jacobian.

The first stage is triangular. Since
\(r_n = p_n + \alpha_n \sigma_n^{2}\) with \(p_n\) and
\(\sigma_n^{2}\) depending only on the previous coefficients, the
diagonal entries of \(\partial r/\partial\alpha\) are \(\sigma_n^{2}\),
and \appref{app:corollaries} gives
\begin{equation}
  \det\!\left(\frac{\partial r}{\partial \alpha}\right)
  =
  \prod_{n=1}^{N} \sigma_n^{2}.
  \label{eq:jac_ra}
\end{equation}
The second stage, \(\alpha \to y\), is component-wise and therefore
diagonal:
\begin{equation}
  \det\!\left(\frac{\partial y}{\partial \alpha}\right)
  =
  \prod_{n=1}^{N} \frac{1}{1 - \alpha_n^{2}}.
  \label{eq:jac_ay}
\end{equation}
The chain rule combines \eqref{eq:jac_ra} and \eqref{eq:jac_ay} into
product forms for the full transformation and its inverse,
\begin{align}
  \left|\det\!\left(\frac{\partial y}{\partial r}\right)\right|
  &=
  \prod_{n=1}^{N}
  \frac{1}{\sigma_n^{2}\,(1 - \alpha_n^{2})},
  \label{eq:jac_ry}\\[3pt]
  \left|\det\!\left(\frac{\partial r}{\partial y}\right)\right|
  &=
  \prod_{n=1}^{N}
  \sigma_n^{2}\,(1 - \alpha_n^{2}).
  \label{eq:jac_yr}
\end{align}
Every factor is a by-product of the Levinson--Durbin recursion of
\sectionref{subsec:cons_bounds}.

Under the change of variables the transformed density reads
\begin{equation}
  p_y(y)
  =
  p_r(r(y))\,
  \left|\det\!\left(\frac{\partial r}{\partial y}\right)\right|,
  \label{eq:cov_transport}
\end{equation}
so residual non-Gaussianity in \(p_y\) can enter both through the
explicit Jacobian factor and through the transformed joint-density
factor \(p_r(r(y))\), which retains any higher-order structure
inherited from the statistics of the underlying field. This is the
density-transport perspective on which the closure test of
\sectionref{subsec:num_gaussianisation} rests.  The same factorisation
transports priors as directly as likelihoods: a product prior on the
individual \(\alpha_n\), each supported on \((-1,1)\), corresponds
through \eqref{eq:jac_ra} to a prior on correlation functions that is
admissible by construction. The link to the general
partial-autocorrelation parametrisation of correlation matrices and
its use for tomographic weak-lensing analyses is discussed in
\sectionref{sec:conclusions}.

\subsection{Volume of the admissible region}
\label{subsec:cons_volume}

Integrating the product-form Jacobian \eqref{eq:jac_ra} over the
open hypercube \((-1,1)^N\) gives the Lebesgue volume of the
admissible region in \(r\)-coordinates. The Jacobian factorises into
single-variable pieces \((1 - \alpha_k^{2})^{N-k}\), so the integral
reduces to a product of Beta integrals, evaluated in
\appref{app:corollaries} in closed form,
\begin{equation}
  V_N =
  2 \prod_{j=1}^{N-1}
  \frac{\sqrt{\pi}\,j!}{\Gamma\!\left(j+\tfrac{3}{2}\right)}.
  \label{eq:volume}
\end{equation}
This is the Toeplitz analogue of the volume formula for general
correlation-matrix sets discussed by \citet{ForresterZhang2020}, and
quantifies the discrepancy between the naive parameter space
\([-1,1]^N\) and the admissible region that actually supports a
Bayesian analysis of correlation functions constrained by a
non-negative one-dimensional power spectrum.

The same admissible region was analysed by
\citet{SchneiderHartlap2009} through a recursive integration of the
interval widths \(\Delta_n = r_{n,\mathrm{u}} - r_{n,\mathrm{l}}\)
(their Eqs.~39--46), carried out with a factor \(1/2\) in each
integration measure. Their result is therefore a fractional volume
relative to the enclosing hypercube \([-1,1]^M\), where their \(M\)
plays the role of our \(N\); SH denote this quantity by \(V_M\) in
their Eq.~(46). To keep it notationally distinct from the Lebesgue
volume \(V_N\) of \eqref{eq:volume}, we shall write it as
\(V_N^{\mathrm{SH}}\) throughout the remainder of this paper, understood
as SH's \(V_M\) evaluated at \(M = N\); the closed form is derived
in \appref{app:corollaries}. The two normalisations are related by
\begin{equation}
  V_N = 2^N \, V_N^{\mathrm{SH}},
  \label{eq:volume_SH}
\end{equation}
so the Levinson--Durbin/PACF derivation and the SH interval
integration compute the same geometric object by different routes and
agree completely. In the SH normalisation, \(V_N^{\mathrm{SH}}\) admits a
direct probabilistic reading as the probability that a correlation
sequence drawn uniformly from \([-1,1]^N\) is admissible, a natural
quantity in prior construction where one asks what fraction of the
naive parameter space is realisable by a positive power spectrum.

\subsection{Boundary behaviour}
\label{subsec:cons_boundary}

The admissible region was coordinatised in
\sectionref{subsec:cons_bijection} by the open hypercube \((-1,1)^N\);
its boundary is reached when some partial autocorrelation
saturates. Under the boundary-index convention of
\sectionref{subsec:proof_degen}, the saturating PACF sits at index
\(m-1\) and \(A_m\) is the first singular Toeplitz matrix. Two aspects
of the boundary must be distinguished carefully. \emph{On the chart
side}, the variance recursion of \sectionref{subsec:cons_bounds} gives
\[
  \sigma_m^{2} = \sigma_{m-1}^{2}\,\bigl(1 - \alpha_{m-1}^{2}\bigr) = 0,
\]
so \(e_0\) is exhausted by its projection onto the intermediate
subspace and the PACF chart -- equivalently, the SH chart \(x_n\) --
ceases to be defined beyond order \(m-1\). \emph{On the sequence
side}, in contrast, the correlation coefficients themselves remain
uniquely determined: \(\det(A_m)=0\) collapses all subsequent SH
intervals to single points, and \citet{Erben2026} shows that, for any
admissible continuation, the coefficients \(r_{m+s}\) are fixed by a
linear recurrence read off from the null vector of \(A_m\). Thus,
wherever the correlation sequence is continued within the
Toeplitz-positive domain, it continues uniquely past the boundary,
even though the chart no longer covers it. The coordinates parametrise
the freedom in the admissible region and terminate once that freedom
is exhausted; any subsequent admissible coefficients are rigidly
determined.

Sequences generated directly in \(\alpha\)-coordinates are admissible
by construction and cannot produce inadmissible draws; measured
correlation estimates, in contrast, may still fall outside the
Toeplitz-positive domain because of estimator noise or catalogue
systematics, as illustrated by the occasional out-of-region
realisations reported by \citet{WilkingRoeselerSchneider2015} for
small random catalogues and as further discussed in the context of the
roundtrip stress test of \sectionref{subsec:num_verification}.

\section{Numerical demonstrations}
\label{sec:num_demonstrations}

This section illustrates the identification \(x_n = \alpha_n\) of
\sectionref{sec:proof} and two consequences of
\sectionref{sec:consequences} through numerical experiments.
\sectionref{subsec:num_geometry} visualises the admissible region at
orders \(N=2\) and \(N=3\) in both \(r\)- and
\(\alpha\)-coordinates. \sectionref{subsec:num_verification} verifies
the mutual consistency of the forward and inverse Levinson--Durbin
recursions on the coordinate
mapping. \sectionref{subsec:num_gaussianisation} applies the identification to
the one-dimensional Gaussian random field of
\citet{SchneiderHartlap2009} and shows that the Fisher
\(z\)-transformation and product-form Jacobian of
\sectionref{subsec:cons_fisher} and \sectionref{subsec:cons_jacobian}
provide an interpretation consistent with the empirical
quasi-Gaussianisation observed by \citet{WilkingSchneider2013} and
\citet{WilkingRoeselerSchneider2015}.

The code implementing the Levinson--Durbin formalism and reproducing
the numerical experiments and figures of this section is publicly
available as the \texttt{ccf} software package \citep{Erben2026ccf}.

\subsection{Geometry of the admissible region}
\label{subsec:num_geometry}

\begin{figure*}[t]
    \centering
    \includegraphics{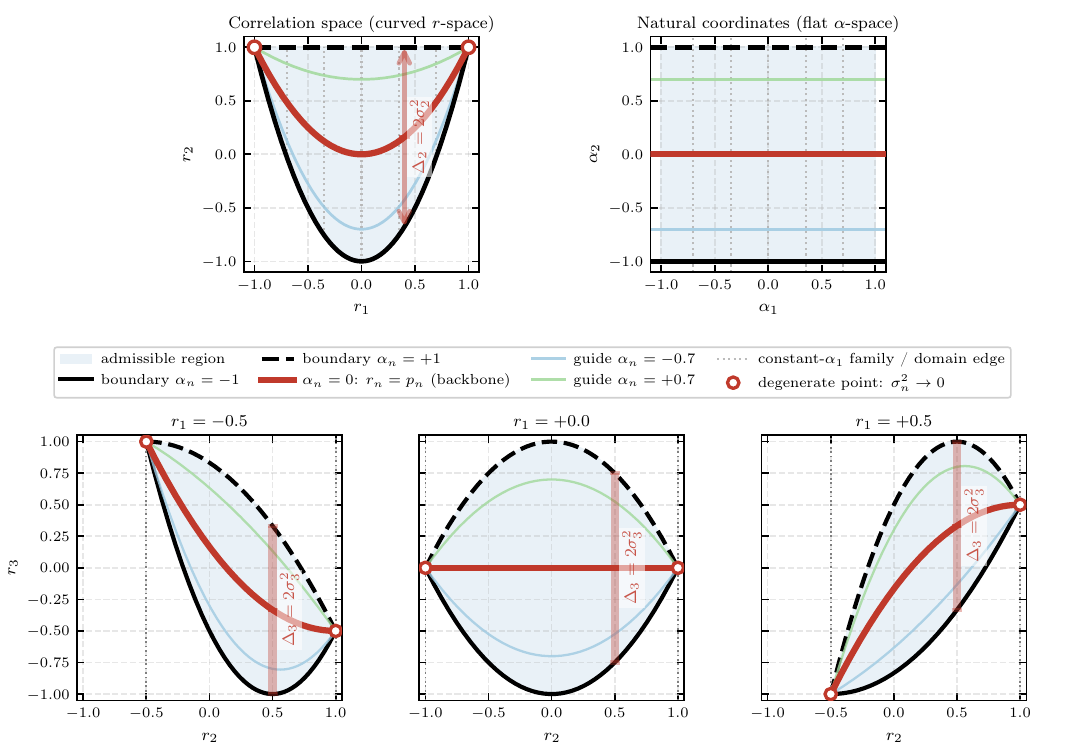}
    \caption{The admissible correlation region shown in correlation
      and natural coordinates, at the first two orders where the
      recursive constraints are non-trivial. \emph{Top row:} the
      two-dimensional case \((r_1, r_2)\), in correlation space (left)
      and natural coordinates (right). The admissible region in
      correlation space is bounded below by the parabola \(r_2 = 2
      r_1^2 - 1\) and above by \(r_2 = +1\); under the change of
      coordinates \((r_1, r_2) \to (\alpha_1, \alpha_2)\) every curve of
      the left panel becomes a straight line and the admissible region
      becomes the open square \((-1, +1)^2\). \emph{Bottom row:} the
      three-dimensional case \((r_1, r_2, r_3)\), visualised through
      three \((r_2, r_3)\)-slices at fixed \(r_1 = -0.5, 0, +0.5\); in
      each slice the shaded region shows the admissible interval of
      \(r_3\) at the given \(r_1\), traced out as \(r_2\) varies.}
    \label{fig:admissible}
\end{figure*}

\figref{fig:admissible} visualises the coordinate change
\(r\to\alpha\) at the first two orders where the recursive constraints
are non-trivial. The red backbones mark the linear predictions
\(p_n\), and the bars indicate the admissible-interval widths
\(\Delta_n = 2\sigma_n^2\) centred on \(p_n\), so the remaining
admissible freedom at each order is measured by the residual variance
\(\sigma_n^2\). The plot makes the projection geometry of the proof
visible: after the predictable component has been projected out,
admissibility reduces to the single residual-correlation bound
\(|\alpha_n| < 1\). The \(N=2\) panel gives the simplest instance of
this straightening; the three-dimensional admissible region at \(N=3\)
is shown through two-dimensional slices at fixed \(r_1\), illustrating
the same recursion one order higher. Higher-order visualisations would
repeat the same local mechanism and are therefore not shown.

\subsection{Roundtrip consistency of the coordinate mapping}
\label{subsec:num_verification}

\begin{figure*}[t]
  \centering
  \includegraphics{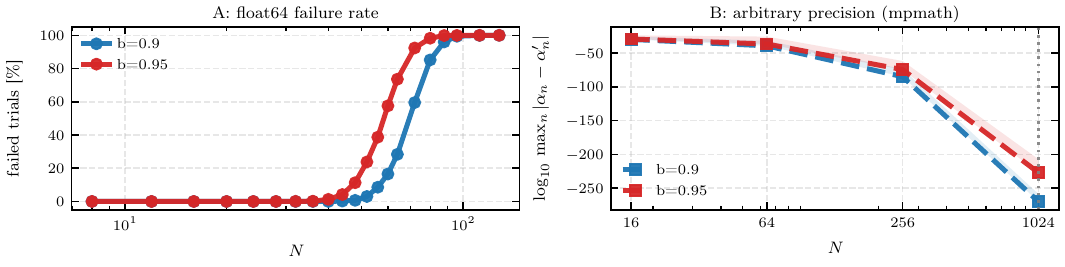}
  \caption{Roundtrip consistency test for the coordinate
    transformation. Blue and red correspond to sampling bounds \(b =
    0.9\) and \(b = 0.95\). \emph{Left:} fraction of float64 runs in
    which the recovered sequence leaves the admissible interval \((-1,
    1)\). \emph{Right:} maximum lag-wise roundtrip error
    \(\log_{10}\max_n |\alpha_n - \alpha'_n|\) in arbitrary
    precision. Dashed curves show medians, shaded bands the
    worst-trial spread; the \(N = 1024\) measurement is a single-point
    sample marked by a dotted vertical.}
  \label{fig:roundtrip}
\end{figure*}

As an internal consistency check for the identified coordinate
mapping, we verify that its two evaluation directions -- the forward
and inverse Levinson--Durbin recursions of
\sectionref{subsec:cons_bounds} -- agree under finite-precision
arithmetic. The experiment isolates properties of the coordinate
mapping itself; questions of detailed numerical optimisation and
implementation strategies lie outside the scope of this
study.

Partial autocorrelations are drawn uniformly from
\(\alpha\in(-b,b)^N\), mapped to correlation coefficients \(r\) via
the forward recursion, and mapped back to \(\alpha'\) via its inverse.
The choices \(b = 0.9\) and \(b = 0.95\) deliberately probe a
near-boundary regime of the hypercube in which the propagated
residual variances \(\sigma_n^{2}\) become very small, so the setup
stresses the coordinate mapping rather than representing typical
applications in the interior of the admissible domain.

The left panel of \figref{fig:roundtrip} shows a sharp float64
failure transition as \(N\) grows: once the residual variances become
too small, the normalisation
\(\alpha_n = (r_n - p_n)/\sigma_n^{2}\) no longer resolves the
residual \(r_n - p_n\) at working precision, and the recovered
\(\alpha'_n\) can leave \((-1, 1)\). The transition marks the point
at which double-precision arithmetic ceases to represent the
near-boundary geometry of the mapping.

The right panel repeats the same experiment in arbitrary precision.
The maximum lag-wise roundtrip error remains at the prescribed precision
level for all tested \(N\), including \(N = 1024\), well beyond the
\(N \sim 50\) angular-bin scale of current cosmic-shear analyses.
Once the arithmetic carries the required dynamic
range,\footnote{The arbitrary-precision arithmetic is provided by
\texttt{mpmath} \citep{mpmath}, with the working precision set to
\(\lceil 0.45\,N \rceil + 24\) decimal digits -- that is,
\(32\), \(53\), \(140\), and \(485\) digits for the four displayed
sample sizes \(N = 16, 64, 256, 1024\) -- calibrated so that the
wider \(b = 0.95\) sampling reaches the roundtrip target of
\(10^{-13}\) with a margin absorbing trial-to-trial variance.} the
forward and inverse recursions of
\sectionref{subsec:cons_bounds} are therefore mutually consistent
on the tested sequences.

\subsection{Quasi-Gaussianisation in natural coordinates}
\label{subsec:num_gaussianisation}
\begin{figure*}[t]
  \centering
  \includegraphics{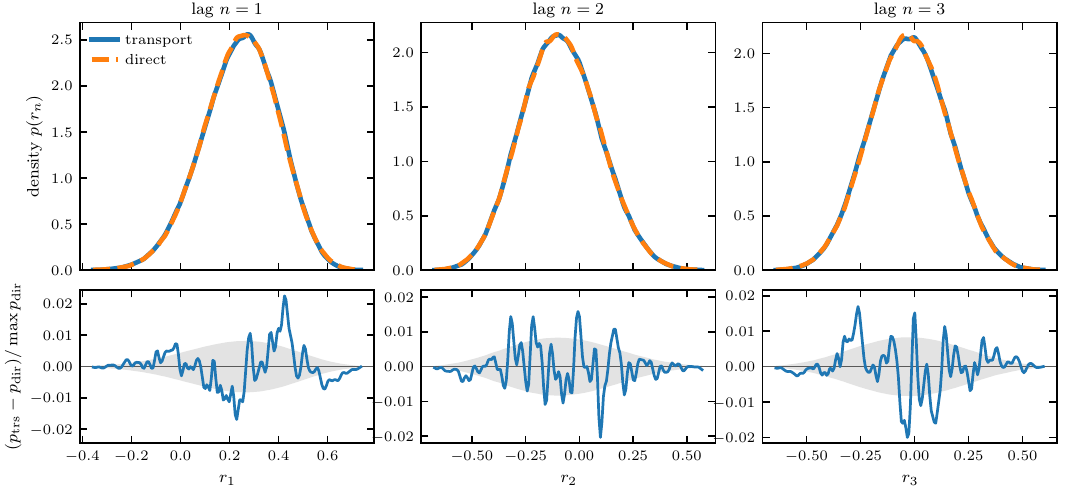}
  \caption{Quasi-Gaussian closure test in natural coordinates.
    \emph{Top row:} marginal densities \(p(r_n)\) for \(n = 1, 2, 3\),
    comparing the direct branch (dashed) with the transport branch
    (solid). \emph{Bottom row:} pointwise density residuals
    \((p_{\mathrm{trs}}-p_{\mathrm{dir}})/\max p_{\mathrm{dir}}\); the
    shaded band shows an approximate pointwise histogram-noise scale,
    obtained by adding the counting-noise variances of the two
    smoothed histograms and neglecting the weak dependence induced by
    fitting the transport model to the direct sample. Numerical
    moments and Kolmogorov--Smirnov distances are listed in
    \tabref{tab:qg_moments}.}
  \label{fig:qg_transport}
\end{figure*}
\begin{table*}
  \renewcommand{\arraystretch}{1.3}%
  \centering
  \caption{Lag-wise Kolmogorov--Smirnov distances
    \(D_{\mathrm{KS}} = \sup_x \left| F_{\mathrm{trs}}(x) - F_{\mathrm{dir}}(x) \right|\)
    between the empirical cumulative distribution functions
    \(F_{\mathrm{dir}}\) and \(F_{\mathrm{trs}}\) of the direct and transport
    branches, Fisher skewnesses \(\gamma = m_3/m_2^{3/2}\), and excess
    kurtoses \(\kappa = m_4/m_2^{2} - 3\), with \(m_k\) the \(k\)-th
    central sample moment. Values refer to the two branches shown in
    \figref{fig:qg_transport} and are given in both \(r\)- and
    \(y\)-coordinates for \(n = 1, 2, 3\).}
  \label{tab:qg_moments}
  \begin{tabular}{c c cc cc cc cc}
    \hline\hline
    \(n\) & \(D_{\mathrm KS}\) &
      \(\gamma_{r}^{\mathrm{dir}}\) & \(\gamma_{r}^{\mathrm{trs}}\) &
      \(\kappa_{r}^{\mathrm{dir}}\) & \(\kappa_{r}^{\mathrm{trs}}\) &
      \(\gamma_{y}^{\mathrm{dir}}\) & \(\gamma_{y}^{\mathrm{trs}}\) &
      \(\kappa_{y}^{\mathrm{dir}}\) & \(\kappa_{y}^{\mathrm{trs}}\) \\
    \hline
    1 & 0.0024 & \(-0.203\) & \(-0.229\) & \(-0.074\) & \(-0.103\) &
                 \(+0.032\) & \(-0.005\) & \(+0.065\) & \(-0.004\) \\
    2 & 0.0013 & \(+0.108\) & \(+0.115\) & \(-0.170\) & \(-0.195\) &
                 \(-0.050\) & \(-0.003\) & \(+0.051\) & \(+0.003\) \\
    3 & 0.0024 & \(+0.001\) & \(+0.018\) & \(-0.192\) & \(-0.219\) &
                 \(-0.002\) & \(-0.003\) & \(+0.049\) & \(-0.015\) \\
    \hline
  \end{tabular}
\end{table*}
As a closure test of the coordinate change \(r\to\alpha\to y\), we ask
whether a Gaussian in \(y\)-space, fitted only to the empirical mean
and covariance of a simulated ensemble, reproduces the direct
\(r\)-marginals of the same ensemble after deterministic transport
back to \(r\). The experiment isolates the marginal-structure content
of a two-moment \(y\)-space surrogate; departures from Gaussianity in
\(y\)-space itself lie outside its scope.

The simulation follows the setup of
\citet[Sect.~2.3]{WilkingSchneider2013}. We generate \(M = 400\,000\)
realisations of a one-dimensional periodic Gaussian random field on
\(N_{\mathrm{grid}} = 32\) grid points with Gaussian-shape power spectrum
\begin{equation}
  \sigma^2(k) \;\propto\; \exp\!\bigl[-(k/k_0)^2\bigr],
  \qquad L k_0 = 80,
  \label{eq:qg_power_spectrum}
\end{equation}
set the highest mode to zero as in \citet{WilkingSchneider2013}, and analyse the normalised
coefficients \(r_1, \ldots, r_{15}\). This ensemble defines the direct
reference; its empirical mean \(\mu_y\) and covariance \(C_y\) in
\(y\)-space are the two-moment input to the transport branch, so the
comparison is a closure test rather than an out-of-sample validation.

For each realisation, the inverse Levinson--Durbin recursion of
\sectionref{subsec:cons_bounds} recovers \(\alpha\) from \(r\), and
\(y_n = \atanh(\alpha_n)\) gives the direct \(y\)-sample. The transport
surrogate
\begin{equation}
  y^{\mathrm{trs}} \sim \mathcal{N}(\mu_y, C_y),
  \label{eq:qg_transport_draw}
\end{equation}
is drawn at the same sample size using the full \(15\)-dimensional
covariance rather than only its diagonal, and \(r^{\mathrm{trs}}\) is
reconstructed by \(\alpha_n^{\mathrm{trs}} = \tanh(y_n^{\mathrm{trs}})\)
followed by the forward recursion. We compare the two branches
through the marginal densities \(p(r_n)\) in \figref{fig:qg_transport}
and the lag-wise Kolmogorov--Smirnov distances, Fisher skewnesses, and
excess kurtoses in \tabref{tab:qg_moments}. This is a
marginal-structure closure test of the coordinate change in
\(r\)-space; the full \(\xi\)-space likelihood of \citet{WilkingSchneider2013}, with its
\(\xi_0\)-marginal and conditioning, lies outside its scope.

The transport branch reproduces the direct-branch \(r\)-marginals at
the few-per-thousand level: the Kolmogorov--Smirnov distances lie
between \(1.3 \times 10^{-3}\) and \(2.4 \times 10^{-3}\), comparable
to the empirical two-sample scale \(\sqrt{2/M} \simeq 2.2 \times
10^{-3}\), and the \(r\)-space Fisher skewnesses and excess kurtoses
of the two branches agree in sign and magnitude at all three displayed
lags, and the residual bands in the lower row of
\figref{fig:qg_transport} confirm the same statement at the density
level. The bulk of the non-Gaussian \(r\)-marginal structure is thus
reproduced by a Gaussian model in \(y\)-space combined with the
deterministic map \(y \to \alpha \to r\). The comparison is not
tautological: the empirical \(y\)-sample is replaced by its two-moment
surrogate, and the residual \(y\)-space skewnesses and excess kurtoses
of the direct branch, of order \(10^{-2}\), quantify the higher-moment
departures of \(p(y)\) that a two-moment model cannot represent.

The conclusion applies within the dynamical range of the
\citet{WilkingSchneider2013} setup used here, where \(\sigma_{y,n} \simeq 0.16\)--\(0.18\) and
\(|\langle y_n\rangle| \leq 0.26\) across the fifteen lags: the
transport probes the moderate rather than the saturation regime of
\(\tanh\). The lag-wise reading given here is the leading-order
picture of a fully multivariate mechanism in which off-diagonal
structure of \(C_y\) propagates into cross-lag structure of \(r\)
through the Levinson--Durbin recursion. More strongly correlated
spectra, for which \(|\alpha_n|\) approaches unity at low lag, would
probe the boundary layer of the same mechanism and are left for future
work.

\section{Conclusions}
\label{sec:conclusions}

We have shown that the coordinate \(x_n\) introduced by
\citet{SchneiderHartlap2009} on the admissible interval of
one-dimensional Toeplitz correlation sequences coincides with the
classical partial autocorrelation coefficient \(\alpha_n\) of the
underlying positive Toeplitz matrix, equivalently with its Schur
reflection coefficient. The result identifies the SH
construction with the classical Verblunsky/Schur parametrisation of
positive Toeplitz matrices, rather than discovering that
parametrisation itself: the mathematical framework had been developed
independently in the time-series and signal-processing literature,
and our contribution is to make this connection explicit.

Under this identification, the specific consequences developed in
\sectionref{sec:consequences} follow as immediate corollaries: the
admissible region becomes the open hypercube \((-1,+1)^N\); the
admissibility bounds and the associated product-form Jacobian for
coordinate transport are supplied by a single Levinson--Durbin forward
pass; and the empirical quasi-Gaussianisation documented by
\citet{WilkingSchneider2013} and \citet{WilkingRoeselerSchneider2015}
acquires a classical reading as the Fisher \(z\)-transformation of the
partial autocorrelations \(\alpha_n\). The one-dimensional
constrained-correlation problem is thereby not solved anew, but
recognised as a specific instance of the classical Toeplitz/PACF
geometry.

Several extensions of the present construction appear promising.
Within the one-dimensional setting, the product-form Jacobian of
\sectionref{subsec:cons_jacobian} connects the SH admissible region
to the partial-autocorrelation parametrisation of general
correlation matrices developed by \citet{LKJ2009}, so that
independent Beta-family priors on the individual \(\alpha_n\) induce
priors on correlation functions that are admissible by construction.
Likewise, the marginal distribution \(p(\xi_0)\) of
\citet{KeitelSchneider2011} and the \(\xi_0\)-conditioning of
\citet{WilkingSchneider2013}, which lie outside the
\(r \to \alpha \to y\) coordinate chain established here, can be
reintroduced on top of the natural coordinates in future
one-dimensional \(\xi\)-space likelihood constructions.

A separate extension concerns vector-valued stationary fields while
retaining the one-dimensional lag ordering: scalar reflection
coefficients are then replaced by matrix-valued coefficients \(K_n\),
whose singular values are restricted to be smaller than unity,
generalising the scalar condition \(|\alpha_n|<1\), and supplied by
classical block-Levinson and matrix-Schur recursions
\citep[Chap.~11]{BrockwellDavis1991}. A potential astronomical setting
is tomographic weak lensing, where several correlated redshift-bin
fields naturally lead to matrix-valued correlation functions. Already
in its scalar form, the present construction likewise remains
applicable to one-dimensional Toeplitz substructures along fixed
directions, angular grids, or separable models. Both settings preserve
the one-dimensional Toeplitz ordering on which the present
construction relies, and should therefore be distinguished from the
problem of genuinely higher-dimensional isotropic correlation
functions, where this ordering is lost.

The extension to genuinely higher-dimensional isotropic correlation
functions is a distinct and more fundamental problem. There, the
one-dimensional Toeplitz-positivity conditions remain necessary but
are no longer sufficient to characterise the admissible correlation
vectors: \citet[Sect.~4]{SchneiderHartlap2009} already showed that
the tightest bounds involve convex hulls of Bessel-function curves.
The ordered Toeplitz structure underlying the PACF construction is
then lost, and the spectral representation instead points naturally
towards a generalised moment problem and the associated convex
geometry of admissible correlation vectors. Rather than seeking a
direct analogue of the one-dimensional reflection coefficients, one
may then ask whether these moment spaces admit natural coordinates
and recursive extensions as additional lags are included.

Building on the constraint framework of
\citet{SchneiderHartlap2009}, the exact distributions of
\citet{KeitelSchneider2011}, and the quasi-Gaussian construction of
\citet{WilkingSchneider2013} with its empirical closure in
\citet{WilkingRoeselerSchneider2015}, the one-dimensional
Toeplitz/PACF case now provides a worked-out reference example: it
exhibits natural coordinates for the admissible
correlation-coefficient region and a benchmark against which
higher-dimensional generalisations can be judged.

\section*{Acknowledgments}

The author thanks Robert Reischke for thoroughly reading the
manuscript and for helpful comments that significantly improved the
article. The author thanks Peter Schneider for making him aware of
this exciting topic while he was looking for examples of research
projects to which computer algebra systems had made a significant
contribution.

\emph{Statement on the use of generative AI tools:}
The author acknowledges the use of OpenAI's ChatGPT, Anthropic's
Claude, and Google's Gemini as interactive assistants during the
development of this work. In particular, these tools were used to
discuss the mathematical structure of the SH
formalism, to learn the concepts underlying partial autocorrelation
functions, and to identify relevant literature, which the author
subsequently retrieved, read, and critically evaluated. The tools
also assisted in drafting and revising parts of the manuscript and
in developing the accompanying code package. All mathematical
arguments, scientific results, analyses, and conclusions were
independently derived or verified by the author, who assumes full
responsibility for the manuscript.

This work is funded by the Deutsche Forschungsgemeinschaft (DFG,
German Research Foundation) under Germany’s Excellence Strategy EXC
3037 – 533607693– Unser dynamisches Universum.

\bibliographystyle{aasjournal}
\bibliography{references}


\begin{appendix}

\section{Derivation of the Levinson--Durbin recursion}
\label{app:ld_derivation}

We derive the update rules \eqref{eq:ld_recursion} of
\sectionref{subsec:cons_bounds} from first principles, in the
geometric language of \sectionref{subsec:proof_non_degen}. All four
updates are consequences of a single identity, namely the orthogonal
decomposition of the prediction space in passing from order \(n\) to
order \(n+1\). We assume \(A_{n+1}\succ0\) throughout; the boundary
case is treated in \sectionref{subsec:cons_boundary}. Classical
references are \citet{Levinson1947,Durbin1960}; for a modern textbook
account see \citet{BrockwellDavis2016}.

We use the notation of \sectionref{subsec:proof_non_degen}, with
\(V_n = \vecspan(e_1,\ldots,e_{n-1})\), endpoint residuals
\(\widetilde e_0^{(n)}\), \(\widetilde e_n^{(n)}\) of common squared
norm \(\sigma_n^{2}\), inner product
\(\langle \widetilde e_0^{(n)}, \widetilde e_n^{(n)}\rangle
   = r_n - p_n = \alpha_n \sigma_n^{2}\), and prediction coefficients
\(\phi_j^{(n)}\) defined by
\(P_{V_n} e_0 = \sum_{j=1}^{n-1} \phi_j^{(n)} e_j\), so that
\(p_n = \sum_{j=1}^{n-1} \phi_j^{(n)} r_{n-j}\). The upper index
\((n)\) denotes the order at which the quantity is evaluated; it is
essential because the derivation involves quantities at orders \(n\)
and \(n+1\) simultaneously.

\paragraph{The orthogonal decomposition of the prediction space.}

Passing from order \(n\) to order \(n+1\) enlarges the intermediate
subspace by one direction, and since \(\widetilde e_n^{(n)}\perp V_n\)
spans that direction, we obtain the orthogonal direct sum
\begin{equation}
  V_{n+1} = V_n \;\oplus\; \vecspan\bigl(\widetilde e_n^{(n)}\bigr).
  \label{eq:app_ld_decomp}
\end{equation}
Any vector \(x\) therefore projects as
\begin{equation}
  P_{V_{n+1}} x = P_{V_n} x
  +
  \frac{\langle x,\, \widetilde e_n^{(n)}\rangle}{\sigma_n^2}
  \,\widetilde e_n^{(n)}.
  \label{eq:app_ld_projection}
\end{equation}
Equation~\eqref{eq:app_ld_projection} is the geometric heart of the
Levinson--Durbin recursion; the update rules below are different
consequences of this single projection identity.

\paragraph{Update for the residual variance.}

We apply \eqref{eq:app_ld_projection} with \(x = e_0\) and subtract
from \(e_0\); using
\(\langle e_0, \widetilde e_n^{(n)}\rangle
   = \langle \widetilde e_0^{(n)}, \widetilde e_n^{(n)}\rangle
   = \alpha_n \sigma_n^{2}\), we obtain the geometric recursion
\begin{equation}
  \widetilde e_0^{(n+1)} =
    \widetilde e_0^{(n)} - \alpha_n\, \widetilde e_n^{(n)}.
  \label{eq:app_ld_e0_update}
\end{equation}
Taking squared norms, and using
\(\|\widetilde e_0^{(n)}\|^2 = \|\widetilde e_n^{(n)}\|^2 = \sigma_n^{2}\),
we find
\begin{equation}
    \sigma_{n+1}^2
    =
    \sigma_n^2\,\bigl(1 - \alpha_n^2\bigr).
  \label{eq:app_ld_sigma_update}
\end{equation}

\paragraph{Update for the prediction coefficients.}

To obtain the update for the coefficients \(\phi_j^{(n)}\), we express
both sides of \eqref{eq:app_ld_e0_update} in the basis
\(e_1,\ldots,e_n\). The left-hand side is
\[
  \widetilde e_0^{(n+1)} = e_0 - \sum_{j=1}^{n} \phi_j^{(n+1)}\, e_j
\]
by definition of the order-\((n+1)\) predictor coefficients; the first
term on the right-hand side is already given in coordinates. For the
second term, we need the coordinate expansion of the endpoint residual
\(\widetilde e_n^{(n)}\).

Denote by \(\psi_j^{(n)}\) the coefficients of the backward predictor,
\(P_{V_n} e_n = \sum_{j=1}^{n-1} \psi_j^{(n)} e_j\). By the Toeplitz
persymmetry \(JBJ = B\) established in
\sectionref{subsec:proof_non_degen}, one has
\(\psi^{(n)} = B^{-1} v = J B^{-1} u = J \phi^{(n)}\), so
componentwise \(\psi_j^{(n)} = \phi_{n-j}^{(n)}\): the backward
predictor is the reversal of the forward predictor. Hence
\[
  \widetilde e_n^{(n)} = e_n - \sum_{j=1}^{n-1} \phi_{n-j}^{(n)}\, e_j.
\]

Substituting both coordinate expansions into
\eqref{eq:app_ld_e0_update} and collecting terms in the basis
\(e_1,\ldots,e_n\), we obtain
\[
  \sum_{j=1}^{n} \phi_j^{(n+1)}\, e_j =
  \sum_{j=1}^{n-1}
     \bigl(\phi_j^{(n)} - \alpha_n\, \phi_{n-j}^{(n)}\bigr)\, e_j
  +
  \alpha_n\, e_n.
\]
Comparing coefficients yields the Levinson--Durbin update
\begin{equation}
    \begin{aligned}
      \phi_j^{(n+1)} &=\phi_j^{(n)} - \alpha_n\, \phi_{n-j}^{(n)},
      \quad j = 1,\ldots,n-1, \\[3pt]
      \phi_n^{(n+1)} &=\alpha_n.
    \end{aligned}
  \label{eq:app_ld_phi_update}
\end{equation}

\paragraph{Prediction and partial autocorrelation at the next order.}

Given the updated coefficients \(\phi_j^{(n+1)}\), the next
predicted correlation is
\begin{equation}
    p_{n+1} = \sum_{j=1}^{n} \phi_j^{(n+1)}\, r_{n+1-j},
  \label{eq:app_ld_p_update}
\end{equation}
and the next partial autocorrelation follows directly from its
definition,
\begin{equation}
    \alpha_{n+1} = \frac{r_{n+1} - p_{n+1}}{\sigma_{n+1}^{2}}\, ,
  \label{eq:app_ld_alpha_update}
\end{equation}
with \(\sigma_{n+1}^2\) obtained from
\eqref{eq:app_ld_sigma_update}. Equations~\eqref{eq:app_ld_sigma_update},
\eqref{eq:app_ld_phi_update}, \eqref{eq:app_ld_p_update}, and
\eqref{eq:app_ld_alpha_update} together close the recursion.

\paragraph{Initial conditions and boundary.}

The recursion starts from \(p_1 = 0\), \(\sigma_1^{2} = 1\), and
\(\alpha_1 = r_1\), as in \sectionref{subsec:ccf_pacf}; at \(n = 1\)
the intermediate subspace \(V_1\) is empty. At the boundary of the
admissible region, saturation of a PACF at some order \(m-1\), i.e.,
\(|\alpha_{m-1}| = 1\), implies \(\sigma_m^{2} = 0\) via
\eqref{eq:app_ld_sigma_update}, so \(e_0 \in V_m\) and the recursion
ceases to define further PACF coordinates; here \(m\) is the boundary
index of \sectionref{subsec:proof_degen} (the index of the first
singular Toeplitz matrix). The corresponding sequence-side behaviour
of the \(r\)-values is developed in \sectionref{subsec:cons_boundary}.

\section{Corollaries of the Levinson--Durbin recursion}
\label{app:corollaries}

In this appendix we collect three consequences of the recursion
derived in \appref{app:ld_derivation}, all of which are used elsewhere
in \sectionref{sec:consequences}. They rest on two ingredients only:
the residual-variance recursion
\(\sigma_{n+1}^2 = \sigma_n^2 (1 - \alpha_n^2)\) established as
\eqref{eq:app_ld_sigma_update}, and the triangular form
\(r_n = p_n + \alpha_n \sigma_n^2\) of the coordinate transformation
\(\alpha \leftrightarrow r\) discussed in
\sectionref{subsec:cons_bijection}. Together they yield a product
formula for the Toeplitz determinant, a product form for the Jacobian
of the transformation, and a closed expression for the volume of the
admissible region.

\paragraph{Product formula for the Toeplitz determinant.}

The interval width \(\Delta_n = 2\sigma_n^{2}\) established in
\sectionref{subsec:proof_non_degen} admits a second expression on the
Toeplitz-determinantal side. \citet{SchneiderHartlap2009} showed
(their Eq.~38, rigorously proved by \citet{Erben2026}) that
\begin{equation}
  \Delta_n = 2\,\frac{\det(A_n)}{\det(A_{n-1})}.
  \label{eq:app_B_delta_det}
\end{equation}
Equating the two expressions for the same interval width yields the
key identity
\begin{equation}
  \sigma_n^{2} = \frac{\det(A_n)}{\det(A_{n-1})},
  \label{eq:app_B_sigma_det}
\end{equation}
which relates the residual variance -- a geometric quantity from the
PACF construction -- to a ratio of Toeplitz determinants. Iterating
\eqref{eq:app_B_sigma_det} from the base case
\(\det(A_1) = \sigma_1^{2} = 1\), and substituting the residual-variance
recursion \(\sigma_k^{2} = \prod_{j=1}^{k-1}(1 - \alpha_j^{2})\)
of \appref{app:ld_derivation}, we obtain the product formula
\begin{equation}
  \det(A_n) = \prod_{k=1}^{n} \sigma_k^{2}
  = \prod_{k=1}^{n-1}\bigl(1 - \alpha_k^{2}\bigr)^{\,n-k},
  \label{eq:app_B_det_product}
\end{equation}
which vanishes precisely when one of the \(\sigma_k^{2}\) does, in
agreement with the degenerate case of \sectionref{subsec:proof_degen}.

\paragraph{The Jacobian of the transformation \(r \leftrightarrow \alpha\).}

The forward recursion \(r_n = p_n + \alpha_n \sigma_n^{2}\), together
with the fact that \(p_n\) and \(\sigma_n^{2}\) depend only on
\(\alpha_1, \ldots, \alpha_{n-1}\), makes the Jacobian matrix
lower-triangular with diagonal entries
\(\partial r_n/\partial \alpha_n = \sigma_n^{2}\). Hence
\begin{equation}
  \det\!\left(\frac{\partial r}{\partial \alpha}\right)
  =
  \prod_{n=1}^{N} \sigma_n^{2}
  =
  \prod_{k=1}^{N-1}\bigl(1 - \alpha_k^{2}\bigr)^{\,N-k},
  \label{eq:app_B_jacobian}
\end{equation}
where the second equality follows from \eqref{eq:app_B_det_product}
applied at order \(N\).

\paragraph{Volume of the admissible region.}

Integrating \eqref{eq:app_B_jacobian} over the open hypercube gives
the volume of the admissible region in \(r\)-space,
\[
\begin{aligned}
  V_N
  &=
  \int_{(-1,1)^{N}}
  \prod_{k=1}^{N-1}
  (1 - \alpha_k^{2})^{\,N-k}\dd^{N}\alpha \\
  &=
  2 \prod_{k=1}^{N-1}
  \int_{-1}^{1}
  (1 - \alpha^{2})^{\,N-k}\dd\alpha ,
\end{aligned}
\]
where the integration over \(\alpha_N\) contributes a factor of 2 and
the integrand factorises into single-variable pieces. Each factor
evaluates through the Beta function to
\[
  \int_{-1}^{1}(1 - \alpha^{2})^{q}\dd\alpha
  =
  \frac{\sqrt{\pi}\,\Gamma(q+1)}{\Gamma\!\left(q + \tfrac{3}{2}\right)}.
\]
Setting \(j = N-k\), we arrive at the closed form
\begin{equation}
  V_N = 2 \prod_{j=1}^{N-1}
  \frac{\sqrt{\pi}\,j!}{\Gamma\!\left(j + \tfrac{3}{2}\right)}.
  \label{eq:app_B_volume}
\end{equation}
A closely related volume formula for general correlation matrices
without the Toeplitz restriction was derived by \citet[see Eqs.~(3.7)
  and (3.8)]{ForresterZhang2020}. Our derivation can be viewed as the
Toeplitz specialisation of their partial-correlation formalism.

Equation~\eqref{eq:app_B_volume} corresponds to Eq.~(46) of
\citet{SchneiderHartlap2009}: in the notation of
\sectionref{subsec:cons_volume}, their fractional volume takes the
closed form
\begin{equation}
  V_M^{\mathrm{SH}} = 2^{\,M(M-1)}\,\prod_{k=2}^{M} B(k,k),
  \label{eq:app_B_VM_SH}
\end{equation}
and symbolic evaluation of \eqref{eq:app_B_volume} and
\eqref{eq:app_B_VM_SH} at \(M = N\) confirms the identity
\(V_N = 2^{\,N}\, V_N^{\mathrm{SH}}\) announced in
\eqref{eq:volume_SH}, with the factor \(2^N\) being precisely the
volume of the enclosing hypercube.

\end{appendix}

\end{document}